\documentclass[aoas,seceqn,nameyear,dvips]{arximspdf}
\usepackage{dcolumn}
\usepackage{graphicx}

\doi{10.1214/09-AOAS298}
\volume{4}
\issue{2}
\pubyear{2010}
\firstpage{1014}
\lastpage{1033}

\makeatletter

\newcolumntype{d}[1]{D{,}{ }{#1}}

\newtheorem{Proposition}{Proposition}

\makeatother

\begin{document}
\begin{frontmatter}

\title{A geometric interpretation of the permutation $\lowercase{p}$-value and its
application in \lowercase{e}QTL studies}
\runtitle{A geometric interpretation of the permutation $p$-value}

\begin{aug}
\author[A]{\fnms{Wei} \snm{Sun}\corref{}\thanksref{t1}\ead[label=e1]{wsun@bios.unc.edu}} and
\author[B]{\fnms{Fred A.} \snm{Wright}\thanksref{t2}\ead[label=e2]{fwright@bios.unc.edu}}
\runauthor{W. Sun and F. A. Wright}
\affiliation{University of North Carolina and University of North Carolina}
\address[A]{Department of Biostatistics\\
Department of Genetics\\
University of North Carolina\\
Chapel Hill, North Carolina\\
USA\\
\printead{e1}}
\address[B]{Department of Biostatistics\\
University of North Carolina\\
Chapel Hill, North Carolina\\
USA\\
\printead{e2}}
\end{aug}

\thankstext{t1}{Supported in part by NHLBI 5 P42 ES05948-16 and 5 P30
ES10126-08.}

\thankstext{t2}{Supported in part by EPA STAR RD832720.}

\received{\smonth{2} \syear{2009}}
\revised{\smonth{9} \syear{2009}}

%
\begin{abstract}
Permutation $p$-values have been widely used to assess the significance
of linkage or association in genetic studies. However, the application
in large-scale studies is hindered by a heavy computational burden. We
propose a geometric interpretation of permutation $p$-values, and based
on this geometric interpretation, we develop an efficient permutation
$p$-value estimation method in the context of regression with binary
predictors. An application to a study of gene expression quantitative
trait loci (eQTL) shows that our method provides reliable estimates of
permutation $p$-values while requiring less than 5\% of the computational
time compared with direct permutations. In fact, our method takes a
constant time to estimate permutation $p$-values, no matter how small the
$p$-value. Our method enables a study of the relationship between nominal
$p$-values and permutation $p$-values in a wide range, and provides a
geometric perspective on the effective number of independent tests.
\end{abstract}

%

\end{frontmatter}

\section{Introduction}\label{sec1}

With the advance of genotyping techniques, high density SNP (single
nucleotide polymorphism) arrays are often used in current genetic
studies. In such situations, test statistics (e.g., LOD scores or
$p$-values) can be evaluated directly at each of the SNPs in order to
map the quantitative/qualitative trait loci. We focus on such
marker-based study in this paper. Given one trait and $p$
markers (e.g., SNPs), in order to assess the statistical significance
of the most extreme test statistic, multiple tests across the
$p$ markers need to be taken into account. In other words, we seek to
evaluate the first step family-wise error rate (FWER), or the
``experiment-wise threshold'' [\citet{Churchill94}]. Because nearby
markers often share similar genotype profiles, the simple Bonferroni
correction is highly conservative. In contrast, the correlation
structure among genotype profiles is preserved across permutations and
thus is incorporated into permutation $p$-value estimation. Therefore,
the permutation $p$-value is less conservative and has been widely used
in genetic studies. Ideally, the \textit{true} permutation $p$-value
can be calculated by enumerating all the possible permutations,
calculating the proportion of the permutations where more extreme test
statistics are observed. In each permutation, the trait is permuted, or
equivalently, the genotype profiles of all the markers are permuted
simultaneously. However, enumeration of the possible permutations is
often computationally infeasible. Permutation $p$-values are often
estimated by randomly permuting the trait a large number of times,
which can still be computationally intensive. For example, to
accurately estimate a permutation $p$-value of 0.01, as many as 1000
permutations may be needed [\citet{Barnard63}, \citet{Marriott79}].

In studies of gene expression quantitative trait loci (eQTL), efficient
permutation $p$-value estimation methods become even more important,
because in addition to the multiple tests across genetic markers,
multiple tests across tens of thousands of gene expression traits need
to be considered [\citet{Kendzioriski06a}, \citet
{Kendzioriski06b}]. One solution
is a two-step procedure, which concerns the most significant eQTL for
each expression trait. First, the permutation $p$-value for the most
significant linkage/association of each expression trait is obtained,
which takes account of the multiple tests across the genotype profiles.
Second, a permutation $p$-value threshold is chosen based on a false
discovery rate (FDR) [\citet{Benjamini95}, \citet{Efron01},
\citet{Storey03}]. This
latter step takes account of the multiple tests across the expression
traits. Following this approach, the computational demand increases
dramatically, not only because there are a large number of expression
traits and genetic markers, but also because stringent permutation
$p$-value threshold, and therefore more permutations must be applied to
achieve the desired FDR. In order to alleviate the computational burden
of permutation tests, many eQTL studies have merged the test statistics
from all the permuted gene expression traits to form a common null
distribution, which, as suggested by empirical studies, may not be
appropriate [\citet{Carlborg05}]. In this paper we estimate the
permutation $p$-value for each gene expression trait separately.

In order to avoid the large number of permutations, some
computationally efficient alternatives have been proposed. \citet
{Nyholt04} proposed to estimate the effective number of independent
genotype profiles (hence the effective number of independent tests) by
eigen-value decomposition of the correlation matrix of all the observed
genotype profiles. Empirical results have shown that, while Nyholt's
procedure can provide an approximation of the permutation $p$-value, it
is not a replacement for permutation testing [\citet
{Salyakina05}]. In
this study we also demonstrate that the effective number of independent
tests is related to the significance level.

Some test statistics (e.g., score test statistics) from multiple tests
asymptotically follow a multivariate normal distribution, and adjusted
$p$-values can be directly calculated [\citet{Conneely07}]. However,
currently at most 1000 tests can be handled simultaneously, due to the
limitation of multivariate normal integration [\citet{GenZ00}].
\citet{Lin05} has proposed to estimate the significance of test
statistics by
simulating them from the asymptotic distribution under the null
hypothesis, while preserving the covariance structure. This approach
can handle a larger number of simultaneous tests efficiently, but it
has not been scaled up to hundreds of thousands of tests, and its
stability and appropriateness of asymptotics have not been validated in
this context.

In this paper we present a geometric interpretation of permutation
$p$-values and a permutation $p$-value estimation method based on this
geometric interpretation. Our estimation method does not rely on any
asymptotic property and, thus, it can be applied when the sample size
is small, or when the distribution of the test statistic is unknown.
The computational cost of our method is constant, regardless of the
significance level. Therefore, we can estimate very small permutation
\mbox{$p$-values}, for example, $10^{-8}$ or less, while estimation by direct
permutations or even by simulation of test statistics may not be
computationally feasible. In principle, our approach can be applied to
the data of association studies as well as linkage studies. However,
the high correlation of test statistics in nearby genomic regions plays
a key role in our approach. Thus, the application to linkage data is
more straightforward. We restrict our discussion to binary genotype
data, which only take two values. Such data include many important
classes of experiments: study of haploid organisms, backcross
populations and recombinant inbred strains. This restriction
simplifies the computation so that an efficient permutation $p$-value
estimation algorithm can be developed. However, the general concept of
our method is applicable to any categorical or numerical genotype data.

The remainder of this paper is organized as follows. In
Section \ref{sec2} we first present the problem setup, followed by an intuitive
interpretation of our method, and finally we describe the more
complicated algebraic details. In Section \ref{sec3} we validate our
method by comparing the estimated permutation $p$-values with the
direct values obtained by a large number of permutations. We also
compare the permutation $p$-values with the nominal $p$-values to
assess the effective number of independent tests. Finally, we discuss
the limitations of our method, and suggest possible improvements.

\section{Methods}\label{sec2}
\subsection{Notation and problem setup}
Suppose there are $p$ markers genotyped in $n$ individuals. The trait
of interest is a vector across the $n$ individuals, denoted by $y =
(y_1,\ldots, y_n)$, where $y_i$ is the trait value of the $i$th
individual. The genotype profile of each marker is also a vector across
the $n$ individuals. Throughout this paper, we use the term ``genotype
profile'' to denote the genotype profile of one marker, instead of the
genotype profile of one individual. Thus, a genotype profile is a point
in the $n$-dimensional space. We denote the entire genotype space as
$\Omega$, which includes $2^n$ distinct genotype profiles.

As mentioned in the \hyperref[sec1]{Introduction}, we restrict our discussion to
binary genotype data, which only take two values. Without loss of
generality, we assume the two values are 0 and 1.
Let $m_1 = (m_{11},\ldots, m_{1n})$ and $m_2 = (m_{21},\ldots, m_{2n})$ be
two genotype profiles. We measure the distance between $m_1$ and $m_2$
by Manhattan distance, that is,
\[
d _{\mathrm{M}} (m_{1} ,m_{2} ) \equiv\sum_{i=1}^{n} |m_{1i}
-m_{2i} | .
\]
We employ Manhattan distance because it is easy to compute and it has
an intuitive explanation: the number of individuals with different
genotypes. In our algorithm the distance measure is only used to group
genotype profiles according to their distances to a point in the
genotype space. Therefore, any distance measure that is a monotone
transformation of Manhattan distance leads to the same grouping of the
genotype profiles, hence the same estimate of the permutation
$p$-value. For binary genotype data, any distance measure $ ( \sum
_{i=1}^{n} |m_{1i} -m_{2i} |^{\tau_1} )^{\tau_2}$ $(\forall\tau_1,
\tau_2 > 0)$ is a monotone transformation of Manhattan distance. We
note, however, this is not true for categorical genotype data with more
than two levels. For example, suppose the genotype of a biallelic
marker is coded by the number of minor allele. Consider three biallelic
markers with genotypes measured in three individuals: $m_1=(0, 0, 0)$,
$m_2=(0, 2, 0)$ and $m_3=(1, 1, 1)$. By Manhattan distance, $d
_{\mathrm{M}} (m_{1} ,m_{2} ) = 2 < d _{\mathrm{M}} (m_{1} ,m_{3} ) =
3$. However, by Euclidean distance, $d  (m_{1} ,m_{2} ) = 2 > d  (m_{1}
,m_{3} ) = \sqrt{3}$. Therefore, different distance measures may not be
equivalent and the optimal distance measure should be the one that is
best correlated with the test-statistic.

In the following discussions we assume one test statistic has been
computed for each marker (locus). Our method can estimate permutation
$p$-value for any test statistic. For the simplicity of presentation,
throughout this paper we assume the test statistic is the nominal
$p$-value.

\subsection{A geometric interpretation of permutation $p$-values}
One fundamental concept of our method is a so-called ``significance
set.'' Let $\alpha$ be a genome-wide threshold used for the collection
of nominal $p$-values from all the markers. A \textit{significance set}
$\Phi(\alpha)$ denotes, for a fixed trait of interest, the set of
possible genotype profiles (whether or not actually observed) with
nominal $p$-values no larger than~$\alpha$. Similarly, we denote such
genotype profiles in the $i$th permutation as $\Phi_i(\alpha)$.
Since permuting the trait is equivalent to permuting all the genotype
profiles simultaneously, $\Phi_i(\alpha)$ is simply a permutation of
$\Phi(\alpha)$.

Whether any nominal $p$-value no larger than $\alpha$ is observed in
the $i$th permutation is equivalent to whether $\Phi_i(\alpha)$
captures at least one observed genotype profile. With this concept of a
significance set, we can introduce the geometric interpretation of the
permutation $p$-value:

\textit{The permutation $p$-value for nominal $p$-value $\alpha$ is, by
definition, the proportion of permutations where at least one nominal
$p$-value is no larger than $\alpha$. This is equivalent to the
proportion of $\{\Phi_i(\alpha)\}$ that capture at least one observed
genotype profile. Therefore, the permutation $p$-value depends on the
distribution of the genotype profiles within $\Phi_i(\alpha)$ and the
distribution of the observed genotype profiles in the entire genotype
space.}

Intuitively, the permutation $p$-value depends on the trait, the
observed genotype profiles and the nominal $p$-value cutoff $\alpha$.
In our geometric interpretation we summarize these inputs by two
distributions: the distribution of all the observed genotype profiles
in the entire genotype space, and the distribution of the genotype
profiles in $\Phi_i(\alpha)$, which include the information from the
trait and the nominal $p$-value cutoff $\alpha$.

\begin{figure}

\includegraphics{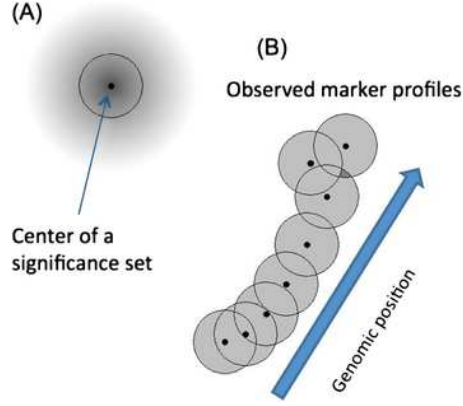}

\caption{A two-dimensional schematic representation of the geometric
interpretation of permutation $p$-value, reflecting genotype profiles
that actually reside in $2^n$-space. \textup{(A)} In the general situation, the
function $P(r, \alpha)$, shown in grayscale, decreases with distance
from the center of a significance set. Under hypersphere assumption,
$P(r, \alpha)$ is either 0 or 1, thus, it can be illustrated by a
hypershpere surrounding the center of the significance set. \textup{(B)} The
space occupied by the series of markers is calculated serially. Denote
the neighborhood region of the $h$th marker as $B_h$. Then the
contribution of the $h$th marker to $\Psi(r_{\alpha})$ is approximated
by $B_h \backslash(B_h \cap B_{h-1})$, where ``$\backslash$'' indicate
set difference. As indicated by the darker shade, this serial counting
approximation is not exact when $(B_h \cap B_k) \protect\notin(B_h \cap
B_{h-1})$, for any $k < h -1$. Note the dot in \textup{(A)} is the center of a
significance set, while the dots in \textup{(B)} are the observed marker
genotype profiles.}
\label{fig1}
\end{figure}

We first consider the genotype profiles in $\Phi_{i}(\alpha)$. For any
reasonably small $\alpha$ (e.g., $\alpha=0.01$), all the genotype
profiles in $\Phi_{i}(\alpha)$ should be correlated, since they are all
correlated with the trait of interest. Therefore, we can imagine these
genotype profiles in $\Phi_i(\alpha)$ are ``close'' to each other in
the genotype space and form a cluster (or two clusters if we separately
consider the genotype profiles positively or negatively correlated with
the trait). In later discussions we show that under some conditions,
the shape of one cluster is approximately a hypersphere in the genotype
space. Then, in order to characterize $\Phi_i(\alpha)$, we need only
know the center and radius of the corresponding hyperspheres. In more
general situations where $\Phi_i(\alpha)$ cannot be approximated by
hyperspheres, we can still define its center and further characterize
the genotype profiles in $\Phi_i(\alpha)$ by a probability
distribution: $P(r, \alpha)$, which is the probability a genotype
profile belongs to $\Phi_i(\alpha)$, given its distance to the center
of $\Phi_i(\alpha)$ is $r$ (Figure \ref{fig1}A). We summarize the information
across all the $\Phi_i(\alpha)$'s to estimate permutation $p$-values.
Since $\{\Phi_{i}(\alpha)\}$ is a one-to-one mapping of all the
permutations, we actually estimate permutation $p$-values by acquiring
all the permutations. Therefore, the computational cost is constant
regardless of $\alpha$. We show this seemingly impossible task is
actually doable. First, because permutation preserves distances among
genotype profiles, the probability distributions from all the
significance sets $\{\Phi(\alpha), \Phi_i(\alpha)\}$ are the same.
Therefore, we only need to calculate it once. Second, the remaining
task is to count the qualifying significance sets, which can be
calculated efficiently using combinations, with some approximations.

%
%

The distribution of the observed genotype profiles in the genotype
space depends on the number of the observed genotype profiles and their
correlation structure. Since $\Phi_i(\alpha)$ may be thought of as
randomly located in the genotype space in each permutation, on average,
the chance that $\Phi_i(\alpha)$ captures at least one observed
genotype profile depends on how much ``space'' the observed genotype
profiles occupy. We argue that such space include the observed genotype
profiles as well as their neighborhood regions. How to define the
neighborhood regions? We first consider the conceptually simple
situation that $\Phi_i(\alpha)$ forms a hypersphere of
radius~$r_{\alpha }$, where the subscript $\alpha$ indicates that $r_{\alpha}$
is a function of $\alpha$. Then $\Phi_i(\alpha)$ captures an observed
genotype profile $m_1$ if its center is within the hypersphere centered
at $m_1$ with radius $r_{\alpha} $. Therefore, the neighborhood region
of $m_1$ is a hypersphere of radius $r_{\alpha} $. We take the union of
the neighborhood regions of all the observed genotype profiles and
denote it by $\Psi(r_{\alpha})$ (Figure \ref{fig1}B). Then we can evaluate
permutation $p$-values by calculating the proportion of significance
sets with their centers within $\Psi(r_{\alpha})$. In the general
situation where the hypersphere assumption does not hold, a
significance set $\Phi_i(\alpha)$ is characterized by a probability
distribution $P(r, \alpha)$. Instead of counting a significance set by
0 or 1, we count the probability it captures at least one observed
genotype profile. We will discuss this estimation method more
rigorously in the following sections.

Before presenting the algebraic details, we emphasize that our method
uses the entire set of the observed genotypes profiles simultaneously.
Specifically, the correlation structure of all the genotype profiles is
incorporated into the construction of $\Psi(r_{\alpha})$. The higher
the correlations between the observed genotype profiles, the more the
corresponding neighborhood regions overlap (Figure \ref{fig1}). This in turn
produces a smaller space $\Psi(r_{\alpha})$, and thus a smaller
permutation $p$-value. In the extreme case when all the observed
genotype profiles are the same, there is effectively only one test and
the permutation $p$-value should be close to the nominal $p$-value.

\subsection{From significance set to best partition}

Explicitly recording all the elements in all the significance sets is
not computationally feasible. We instead characterize each significance
set by a best partition, which can be understood as the center of the
significance set, and a probability distribution: the probability that
one genotype profile belongs to the significance set, given its
distance to the best partition.

We first define best partition. The \textit{best partition} for $\Phi
(\alpha)$ [or $\Phi_i(\alpha)$] is a partition of the samples that is
most significantly associated with the trait (or the $i$th
permutation of the trait). For a binary trait, the trait itself
provides the best partition. For a quantitative trait, we generate the
best partition by assigning the smallest $t$-values to one phenotype
class and the other $(n-t)$-values to another phenotype class. We
typically use $t = n/2$ as a robust choice. The robustness of this
choice is illustrated by the empirical evidence in the Supplementary
Materials [\citet{Sun09}]. Given $t$, we refer to all the
possible best
partitions (partitions that divide the $n$ individuals into two
groups of size $t$ and $n-t$) as \textit{desired partitions}.
The total number of distinct desired partitions, denoted by $N_p$, is
%
%
\begin{equation} \label{3np}
N_{p} =\cases{
\pmatrix{n \cr t}, &\quad if $t\ne n/2$,\vspace*{2pt}\cr
\dfrac{1}{2} \pmatrix{n \cr t}, &\quad if $t=n/2$.}
\end{equation}
When $t=n/2$, there are ${n \choose t}$ ways to choose $t$ individuals,
but two such choices correspond to one partition, that is why we need
the factor $1/2$. For a binary trait, the desired partitions and the
significance sets have one-to-one correspondence and, thus, $N_p$ is
the total number of significance sets (or the total number of
permutations). For a quantitative trait, $N_p$ is much smaller than the
total number of significance sets. In fact, each desired partition
corresponds to $t!(n-t)!$ distinct significance sets (or permutations).
Since we restrict our study for binary genotype, this definition of
best partition can be understood as the projection of the trait into
the genotype space. This projection is necessary to utilize the
geometric interpretation of permutation $p$-value. Note the best
partition does not replace the trait since the trait data is still used
in calculating $P(r, \alpha)$. The projection of trait into genotype
space is less straightforward when the genotype has three or more
levels, though it is still feasible. Further theoretical and empirical
studies are needed for such genotype data.

Next, we study the probability that one genotype profile belongs to a
significance set given its distance to the best partition of the
significance set. Each desired partition, denoted as $\mathit{DP}_j$, has
perfect correspondence with two genotype profiles, depending on whether
the first $t$-values are 0 or 1. We denote these two genotype profiles
as $m_{j}^0$ and $m_{j}^1$, respectively. The distance between one
genotype profile $m_1$ and one desired partition $\mathit{DP}_j$ is defined as
\[
d_{\mathrm{M}} (m_{1} ,\mathit{DP}_{j} ) \equiv\min
_{a=0,1} \{d_{\mathrm{M}} (m_{1} ,m_{j}^{a} )\} .
\]
Suppose $\mathit{DP}_j$ is the best partition of the significance set $\Phi
_i(\alpha)$. In general, the smaller the distance from a genotype
profile to $\mathit{DP}_j$, the greater the chance it falls into $\Phi_i(\alpha
)$. Thus, the genotype profiles in $\Phi_i(\alpha)$ form two clusters,
centered on $m_{j}^0$ and $m_{j}^1$, respectively. The probability
distribution we are interested in is
\[
\Pr\bigl(m_{1} \in\Phi_i(\alpha)
|\forall m_{1} \in\Omega, d _{\mathrm{M}} (m_{1} ,\mathit{DP}_{j}
)=r \bigr) .
\]
This probability certainly depends on the trait $y$. However,
because all of our inference is conducted on $y$, we have suppressed
$y$ in the notation. A similar probability distribution can be defined
for the significance set $\Phi(\alpha)$. Because the permutation-based
mapping $\Phi (\alpha) \rightarrow\Phi_i(\alpha)$ preserves distances,
the distributions for $\Phi(\alpha)$ and $\Phi_i(\alpha)$ are the same
and, thus, we need only quantify the distribution for $\Phi(\alpha)$.
We denote the best partition of the unpermuted trait $y$ as
$\mathit{DP}_y$, and denote the two genotype profiles corresponding to
$\mathit{DP}_y$ as $m_{y}^0$ and $m_{y}^1$, then we define the
distribution as follows:
%
%
\begin{equation} \label{equ1}
P(r,\alpha) \equiv\Pr\bigl(m_{1} \in\Phi(\alpha)
|\forall m_{1} \in\Omega, d _{\mathrm{M}} (m_{1} ,\mathit{DP}_{y}
)=r \bigr) .
\end{equation}
Let
%
%
\begin{equation} \label{equ2}
P(m_{y}^{a} ,r, \alpha) \equiv\Pr\bigl(m_{1} \in\Phi(\alpha)
|{\forall m_{1} \in\Omega, d}_{\mathrm{M}} (m_{1} ,m_{y}^{a}
)=r \bigr),
\end{equation}
where $a = 0, 1$. We have the following conclusion.
\begin{Proposition}\label{Proposition1}
$P(r, \alpha) = P(m_{y}^0, r,
\alpha) = P(m_{y}^1, r, \alpha)$ for any $r< n/2$.
\end{Proposition}

The proof is in the Supplementary Materials [\citet{Sun09}].

By Proposition \ref{Proposition1}, in order to estimate $P(r, \alpha)$, we can simply
estimate $P(m_{y}^0, r, \alpha)$. Specifically, we first randomly
generate $H$ genotype profiles $\{m_h\dvtx h = 1,\ldots, H\}$ so that $d _{\mathrm{M}}
(m_h, m_{y}^0) = r$. To generate $m_h$, we flip the
genotype of $m_{y}^0$ for $r$ randomly chosen individuals. Then
$P(r, \alpha)$ is estimated by the proportion of \{$m_h$\} that yield
nominal $p$-values no larger than $\alpha$.

In summary, we characterize a significance set $\Phi_i(\alpha)$ by the
corresponding best partition and the probability distribution $P(r,
\alpha)$. All the distinct best partitions are collectively referred to
as desired partitions. This characterization of significance sets has
two advantages. First, the probability distribution $P(r, \alpha)$ is
the same across all the significance sets, so we need only calculate it
once. This is because the probability distribution relies on distance
measure, which is preserved across significance sets (permutations).
Second, for a quantitative trait, one desired partition corresponds to
a large number of significance sets; therefore, we significantly reduce
the dimension of the problem by considering desired partitions instead
of significance sets.

\subsection{Estimating permutation $p$-values under a hypersphere assumption}
By the definition of a significance set, we can calculate the
permutation $p$-value by counting the number of significance sets that
capture at least one observed genotype profile. However, it is still
computationally infeasible to examine all significance sets. Therefore,
in the previous section we discuss how to summarize the significance
sets by desired partitions and a common probability distribution. In
this and the next sections, we study how to estimate permutation
$p$-values by ``counting'' desired partitions.

To better explain the technical details, we begin with a simplified
situation, by assuming there is an $r_\alpha$ such that $P(r, \alpha)$
= 1 if $r \leq r_{\alpha}$ and $P(r, \alpha)$ = 0 otherwise. This is
equivalent to assuming $\Phi(\alpha)$ or $\Phi_{i}(\alpha)$ occupies
two hyperspheres with radius $r_{\alpha}$. This \textit{hypersphere
assumption} turns out to be a reasonable approximation for a balanced
binary trait (see Supplementary Materials [\citet{Sun09}]).

Let $\{m_{o,k}, 1\leq k \leq p\}$ be the observed $p$ genotype
profiles. We formally define the space occupied by the observed
genotype profiles and their neighborhood regions as
\[
\Psi(r_{\alpha} ) \equiv\Bigl\{ m_1\dvtx m_1\in\Omega, \min_{1\le k\le p}
\{d _{\mathrm{M}}(m_1, m_{o,k})\} \le r_{\alpha} \Bigr\},
\]
that is, all the possible genotype profiles within a fixed distance
$r_{\alpha}$ from at least one of the observed genotype profiles. We
have the following conclusion under the hypersphere assumption.
\begin{Proposition}\label{Proposition2}
Consider a significance set $\Phi_i(\alpha)$ occupying two hyperspheres
centered at $m_{j}^0$ and $m_{j}^1$, respectively,\vspace*{1pt} with radius
$r_{\alpha}$. $\Phi_i(\alpha)$ corresponds to one permutation of the
trait. The minimum nominal $p$-value of this permutation is no larger
than $\alpha $ iff at least one of $m_{j}^0$ and $m_{j}^1$ is within
$\Psi (r_{\alpha})$.
\end{Proposition}

The proof is in the Supplementary Materials [\citet{Sun09}].

Based on Proposition \ref{Proposition2}, we can calculate the
permutation $p$-value by counting the number of significance sets with
at least one of its centers belonging to~$\Psi(r_{\alpha} )$. Note
under this hypersphere assumption, for any fixed $\alpha$ (hence fixed
$r_{\alpha}$), the significance sets are completely determined by the
centers of the corresponding hyperspheres. Thus, there is a one-to-one
mapping between significance sets and their centers, the desired
partitions. Counting significance sets is equivalent to counting
desired partitions. Therefore, we can estimate the permutation
$p$-value by counting the number of desired partitions. Specifically,
let the distances from all the observed genotype profiles to
$\mathit{DP}_j$, sorted in ascending order, be $(r_{j1}, \ldots,
r_{jp})$. Then under the hypersphere assumption, the permutation
$p$-value for significance level $\alpha$ is
%
%
\begin{equation} \label{eq:04}
|\{ \mathit{DP}_{j} \dvtx r_{j1} \le r_{\alpha} \} |/N_{p} \equiv C(r_\alpha)/N_{p},
\end{equation}
where $N_p$ is the total number of desired partitions, and $C(r_\alpha)
\equiv|\{ \mathit{DP}_{j} \dvtx r_{j1} \le r_{\alpha} \} |$ is the number of desired
partitions within a fixed distance $r_\alpha$ from at least one of the
observed genotype profiles. The calculation of $C(r_ \alpha)$ will be
discussed in the next section.

We note that the hypersphere assumption is not perfect even for the
balanced binary trait. We employ the hypersphere assumption to give a
more intuitive explanation of our method. In the actual implementation
of our method, even for a balanced binary trait, we still use the
general approach to estimate permutation $p$-values, as described in
the next section.

\subsection{Estimating permutation $p$-values in general situations}

In general situations where the hypersphere assumption does not hold,
we estimate the permutation $p$-value by
%
%
\begin{equation} \label{eq:05}
\sum_{j} \Pr(\mathit{DP}_{j} ,\alpha) /N_{p} ,
\end{equation}
where $\Pr(\mathit{DP}_j, \alpha)$ is the probability that the minimum nominal
$p$-value $\leq$ $\alpha$ given $\mathit{DP}_j$ is the best partition. Equation
(\ref{eq:05}) is a natural extension of equation (\ref{eq:04}) by
replacing the counts with the summation of probabilities. It is worth
noting that in the previous section, one desired partition corresponds
to one significance set given the hypersphere assumption. However, in
general situations, one desired partition may correspond to many
significance sets. Therefore, $\Pr(\mathit{DP}_j, \alpha)$ is the average
probability that the minimum nominal $p$-value $\leq$ $\alpha$ for all
the significance sets centered at $\mathit{DP}_j$. Taking averages does not
introduce any bias to permutation $p$-value estimation, because
permutation $p$-value is itself an average. Here we just take the
average in two steps. First, we average across all the significance
sets (or permutations) corresponding to the same desired partition to
estimate $\Pr(\mathit{DP}_j, \alpha)$. Second, we average across desired
partitions.

Let all the desired partitions whose distances to an observed genotype
profile $m_{o,k}$ are no larger than $r$ be $B_k(r)$, that is,
\[
B_k(r) \equiv\{ \mathit{DP}_j\dvtx d _{\mathrm{M}}(m_{o,k}, \mathit{DP}_j) \leq r\},
\]
where $1 \leq k \leq p$. Assume the observed genotype profiles
$\{m_{o,k}\}$ are ordered by the chromosomal locations of the
corresponding markers. We employ the following two approximations to
estimate $\sum_{j}\Pr(\mathit{DP}_{j} , \alpha) $:

\begin{enumerate}
\item\textit{shortest distance approximation}:
\[
\Pr(\mathit{DP}_{j} ,\alpha)\approx P(r_{j1} ,\alpha),
\]

\item\textit{serial counting approximation}:
\[
C(r)\approx C_{U} (r)\equiv\sum_{h=1}^{p} |B_{h} (r) |-
\sum_{h=2}^{p} |B_{h}(r)\cap B_{h-1}(r) |,
\]
\end{enumerate}
where $C(r)$ has been defined in equation (\ref{eq:04}).
\begin{Proposition}\label{Proposition3}
As long as $\alpha$ is reasonably small, for example, $\alpha< 0.05$,
there exist $r_{L} <r_{U}$, such that $P(r,\alpha)=1$, if $r\le r_{L}
$; $P(r,\alpha)=0$, if $r\ge r_{U} $. Given the shortest distance and
the serial counting approximations,
%
%
\begin{eqnarray} \label{eq:06}
\sum_{j}\Pr(\mathit{DP}_{j} ,\alpha) &\approx& \sum_{j}P(r_{j1} ,\alpha)
\nonumber\\[-8pt]\\[-8pt]
&\approx& C_{U} (r_{L} )+\sum_{r=r_{L} +1}^{r_{U} -1}
\bigl[P(r,\alpha) \bigl(C_{U} (r)-C_{U} (r-1) \bigr) \bigr].\nonumber
\end{eqnarray}
When $\alpha$ is extremely small, for example, $\alpha=10^{-20}$, it is
possible $r_L=0$. We define $C_{U} (0)=0$ to incorporate this situation
into equation (\ref{eq:06}).
\end{Proposition}

In the Supplementary Materials [\citet{Sun09}], we present the
derivation of Proposition \ref{Proposition3}, as well as Propositions
4 and 5 that provide the algorithms
to calculate $|B_h(r)|$ and $|B_h(r) \cap B_{h-1}(r)|$, respectively.
Therefore, by Propositions~\ref{Proposition3}--5, we
can estimate the permutation $p$-value by equation (\ref{eq:05}).

The rationale of shortest distance approximation is as follows. If the
space occupied by a significance set is approximately two hyperspheres,
this approximation is exact. Otherwise, if $\alpha$ is small, which is
the situation where direct permutation is computationally unfavorable,
this approximation still tends to be accurate. This is because when
$\alpha$ is smaller, the genotype profiles within the significance set
are more similar and, hence, the significance set is better
approximated by two hyperspheres. In Section \ref{sec3} we report
extensive simulations to evaluate this approximation.

The serial counting approximation can be justified by the property of
genotype profiles from linkage data, and (with less accuracy) in some
kinds of association data. In linkage studies, the similarity between
genotype profiles is closely related to the physical distances, with
conditional independence of genotypes between loci given the genotype
at an intermediate locus. Therefore, the majority of the points in
$B_{h}(r)\cap B_{h-k}(r)$ ($2 \leq k \leq h-1$) are already included in
$B_{h}(r)\cap B_{h-1}(r)$ (Figure~\ref{fig1}B) and, thus,
\[
B_{h}(r)\cap\biggl(\bigcup_{1 \leq k \leq h-1} B_k(r)\biggr) \approx
B_{h}(r)\cap B_{h-1}(r).
\]
Then, we have
\begin{eqnarray*}
C(r) &=& \sum_{k=1}^p |B_{k}(r) | - \sum_{h=2}^p
\biggl|B_{h}(r)\cap\biggl(\bigcup_{1 \leq k \leq h-1} B_k(r)\biggr) \biggr| \\
&\approx& \sum_{k=1}^p |B_{h}(r) | - \sum_{h=2}^p
|B_{h}(r)\cap B_{h-1}(r) |.
\end{eqnarray*}

Our method has been implemented in an R package named permute.t, which
can be downloaded from \href{http://www.bios.unc.edu/\textasciitilde wsun/software.htm}{http://www.bios.unc.edu/\textasciitilde wsun/software.htm}.

\section{Results}\label{sec3}

\subsection{Data}

\setcounter{footnote}{2}

We analyzed an eQTL data set of 112 yeast segregants generated from two
parent strains [\citet{Brem05a}, \citet{Brem05b}].
Expression levels of 6229
genes and genotypes of 2956 SNPs were measured in each of the
segregants. Yeast is a haploid organism and, thus, the genotype profile
of each marker is a binary vector of 0's and 1's, indicating the parental
strain from which the allele is inherited. We dropped 15 SNPs that had
more than 10\% missing values, and then imputed the missing values in
the remaining SNPs using the function fill.geno in R/qtl [\citet
{Broman03}]. Finally, we combined the SNPs that have the same genotype
profiles, resulting in 1017 distinct genotype profiles.\footnote{Most
SNPs sharing the same genotype profiles are adjacent to each other,
although there are 10~exceptions in which the SNPs with identical
profiles are separated by a few other SNPs. In all the 10 exceptions,
the gaps between the identical SNPs are less than 10 kb. We recorded the
position of each combined genotype profile as the average of the
corresponding SNPs' positions.} As expected, genotype profiles between
chromosomes have little correlation (Figure \ref{fig2} in the Supplementary
Materials [\citet{Sun09}]), while the correlations of genotype profiles
within one chromosome are closely related to their physical proximity
(Figure \ref{fig3} in the Supplementary Materials [\citet{Sun09}]).

\subsection{Evaluation of the shortest distance approximation}

We evaluate the shortest distance approximation $\Pr(\mathit{DP}_{j}
,\alpha )\approx P(r_{j1} ,\alpha)$ in this section. Because the
permutation $p$-value is actually estimated by the average of\break
$\Pr(\mathit{DP}_{j} ,\alpha )$ [equation~(\ref{eq:05})], it is
sufficient to study the average of $\Pr(\mathit{DP}_{j} ,\alpha)$
across all the $\mathit{DP}_j$'s having the same $r_{j1}$. Specifically,
we simulated 50 desired partitions $\{\mathit{DP}_j, j = 1,\ldots,
50\}$ such that, for each $\mathit{DP}_j$, $r_{j1} = r$. Suppose
$\mathit{DP}_j$ divides the $n$ individuals into two groups of size $t$
and $n-t$; then $\mathit{DP}_j$ is consistent with $t!(n-t)!$
permutations of the trait. We randomly sampled 1000 such permutations
to estimate $\Pr (\mathit{DP}_{j} ,\alpha)$. We then took the average
of these 50 $\Pr(\mathit{DP}_{j} ,\alpha)$'s, denoted it as
$\bar{\rho}(r)$, and compared it with $P(r, \alpha)$.

\begin{figure}

\includegraphics{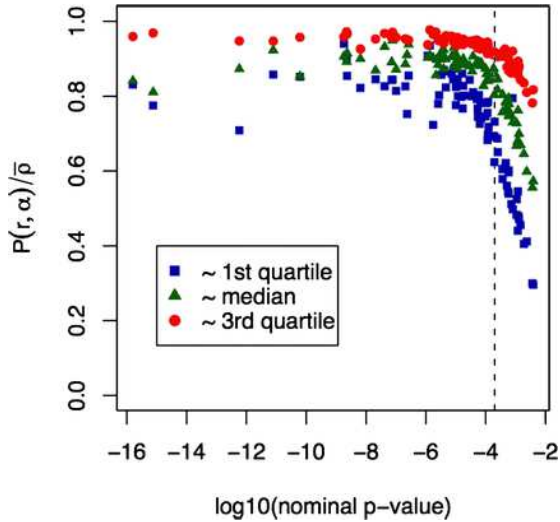}

\caption{Evaluation of the shortest distance approximation using 88
randomly selected gene expression traits. For each gene expression
trait, the ratio $P(r, \alpha)/\bar{\rho}(r)$ is plotted at three r's,
which are approximately the 1st quartile, median and 3rd quartile of
$P(r, \alpha)$ when $P(r, \alpha)$ is between 0 and 1. The vertical
broken line indicates the nominal $p$-value $2\times10^{-4}$, which
corresponds to genome-wide permutation $p$-value $0.05\sim0.10$.}
\label{fig2}
\end{figure}

We randomly selected 88 gene expression traits. For each gene
expression trait, we chose $\alpha$ to be the smallest nominal
$p$-value (from $t$-tests) across all the 1,107 genotype profiles. We
first estimated $P(r, \alpha)$ and $\bar{\rho}(r)$, and then examined
the ratio $P(r, \alpha)/\bar{\rho}(r)$ at three distances $r_{i} $,
$i=1, 2, 3$, where $r_i = \arg\min_r \{|P(r, \alpha) -
0.25i|\}$,
that is, the approximate 1st quartile, median and 3rd quartile of $P(r,
\alpha)$ when $P(r, \alpha)$ is between 0 and 1 (Figure \ref{fig2}). For the
%
%
genes with larger nominal $p$-values, $P(r, \alpha)/\bar{\rho}(r)$ can
be as small as 0.4. Thus, the shortest distance approximation is
inaccurate. We suggest estimating the permutation $p$-values for the
genes with larger nominal $p$-values by a small number of direct
permutations, although, in practice, such nonsignificant genes may be
of little interest. After excluding genes with nominal $p$-values
larger than $2\times10^{-4}$, on average, $P(r, \alpha)/\bar{\rho}(r)$
is 0.80, 0.88, 0.95 for the 1st, 2nd and 3rd quartile respectively. We
chose the threshold $2\times10^{-4}$ because it approximately
corresponds to permutation $p$-value $0.05\sim0.10$ (see Section 3.4.
Comparing permutation $p$-value and nominal $p$-value). It is worth
emphasizing that when we estimate permutation \mbox{$p$-values}, we average
across $\mathit{DP}_{j}$'s. In many cases, $P(r_{j1} ,\alpha) = 0$ or 1 and,
thus, $\Pr(\mathit{DP}_{j} ,\alpha) = P(r_{j1} ,\alpha)$. Therefore, after
taking the average across $\mathit{DP}_{j}$'s, the effects of those cases with
small $P(r, \alpha)/\bar{\rho}(r)$ will be minimized.

\subsection{Permutation $p$-value estimation for a balanced binary
trait---evaluation of the serial counting approximation}

Using the genotype data from the yeast eQTL data set, we performed a
genome-wide scan of a simulated balanced binary trait, with 56 0's and
56 1's. The standard chi-square statistic was used to quantify the
linkages. As we discussed before, for a balanced binary trait, the
space occupied by a significance set is approximately two hyperspheres,
and the shortest distance approximation is justified. This conclusion
can also be validated empirically by examining $P(r, \alpha)$. As shown
in Table 3 of the Supplementary Materials [\citet{Sun09}], for each
$\alpha$, there is an $r_{\alpha}$, such that $P(r, \alpha)=1$ if $r
\le r_{\alpha}$, and $P(r,\alpha) \approx0$ if $r > r_\alpha$. From
the sharpness of the boundary we can see that a significance set indeed
can be well approximated by two hyperspheres. Given that the shortest
distance approximation is justified, we can evaluate the accuracy of
the serial counting approximation by examining the accuracy of
permutation $p$-value estimates.

%
%
\begin{table}
\caption{Comparison of permutation $p$-value estimates for a balanced
binary trait. Values at the column of ``Permutation $p$-value'' are
estimated via 500,000 permutations. Values at the columns ``Permutation
$p$-value estimate I/II'' are estimated by our method before and after
perturbing~the~locations of the SNPs}\label{table1}
\begin{tabular*}{\tablewidth}{@{\extracolsep{\fill}}ld{1.10}d{1.10}d{1.10}@{}}
\hline
\textbf{Nominal} & \multicolumn{1}{c}{\textbf{Permutation}} & \multicolumn{1}{c}{\textbf{Permutation}}
& \multicolumn{1}{c@{}}{\textbf{Permutation}} \\
\textbf{$\bolds p$-value} & \multicolumn{1}{c}{\textbf{$\bolds p$-value}}
& \multicolumn{1}{c}{\textbf{$\bolds p$-value}} & \multicolumn{1}{c@{}}{\textbf{$\bolds p$-value}}\\
\textbf{cutoff} &  & \multicolumn{1}{c}{\textbf{estimate I}}
& \multicolumn{1}{c@{}}{\textbf{estimate II}}\\
\hline
$10^{-3}$ & ,0.19 & ,0.21 & ,0.41 \\
$10^{-4}$ & ,0.02 & ,0.021 & ,0.039 \\
$10^{-5}$ & ,\mbox{$2.0\times10^{-3}$} & ,\mbox{$1.9\times10^{-3}$}
& ,\mbox{$2.9\times10^{-3}$} \\
$10^{-6}$ & ,\mbox{$2.4\times10^{-4}$} & ,\mbox{$2.2\times10^{-4}$}
& ,\mbox{$3.1\times10^{-4}$} \\
\hline
\end{tabular*}
\end{table}

%
%
\begin{figure}[b]

\includegraphics{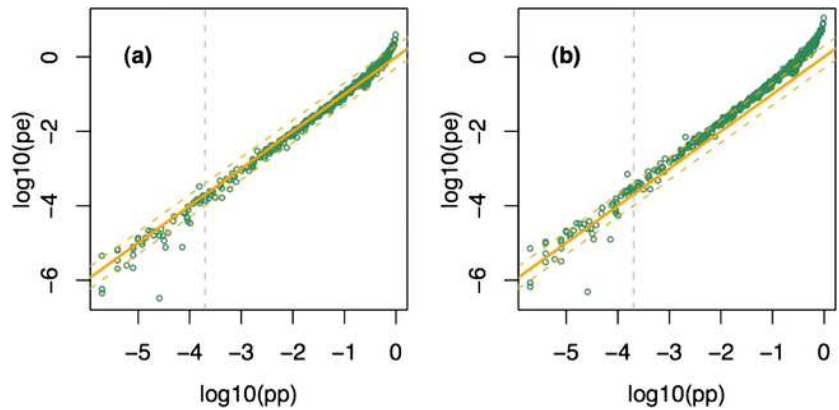}

\caption{Comparison of permutation $p$-values estimated by our method
(denoted as pe) or by direct permutations (denoted as pp) for 500
randomly selected gene expression traits (each gene corresponds to one
point in the plot). \textup{(a)} Using the original genotype data. \textup{(b)} Using the
location-perturbed genotype data. Each gene expression trait is
permuted up to 500,000 times to estimate pp. Thus, the smallest
permutation $p$-value is $2\times10^{-6}$, and we have more confidence
for those permutation $p$-values bigger than $2\times10^{-4}$
(indicated by the vertical line). The degree of closeness of the points
to the solid line ($y = x$) indicates the degree of consistency of the
two methods. The two broken lines along the solid line are $y=x \pm
\log _{10}(2)$ respectively, which, in the original $p$-value scale,
are pe${}={}$0.5pp and pe${}={}$2pp, respectively.}\label{fig3}
\end{figure}

The accuracy of the serial counting approximation relies on the
assumption that the adjacent genotype profiles are more similar than
the distant ones. We dramatically violate this assumption by randomly
ordering the SNPs in the yeast eQTL data. As shown in Table \ref{table1}, the
permutation $p$-value estimates from the original genotype data are
close to the permutation $p$-values estimated by direct permutations,
whereas the estimates from the location-perturbed genotype data are
systematically biased.

\subsection{Permutation $p$-value estimation for quantitative traits}

We randomly selected 500 gene expression traits to evaluate our
permutation $p$-value estimation method in a systematic manner. We used
$t$-tests to evaluate the linkages between gene expression traits and
binary markers. For each gene expression trait, we first identified the
genome-wide smallest $p$-value, and then estimated the corresponding
permutation $p$-value by either our method or by direct permutations
[Figure~\ref{fig3}(a)]. For those relatively larger permutation $p$-values
($>$0.1), the estimates from our method tend to be inflated. Some of them
are even greater than 1. This is because the serial counting
approximation is too loose for larger permutation \mbox{$p$-values}, due to
the fact that each significance set occupies a relatively large space.
Nevertheless, the two estimation methods give consistent results for
those permutation $p$-values smaller than 0.1. We also estimated the
permutation $p$-values after perturbing the order of the SNPs [Figure
\ref{fig3}(b)]. As expected, the permutation \mbox{$p$-value}
estimates are inflated.

The advantage of our method is the improved computational efficiency.
The computational burden of our method is constant no matter how small
the permutation $p$-value is. To make a fair comparison, both our
estimation method and direct permutation were implemented in C. In
addition, for direct permutations, we carried out different number of
permutations for different gene expression traits so that a large
number of permutations were performed only if they were needed.
Specifically, we permuted a gene expression trait 100, 1000, 5000,
10,000, 50,000 and 100,000 times if we had 99.99\% confidence that the
permutation $p$-value of this gene was bigger than 0.1, 0.05, 0.02,
0.01, 0.002 and 0.001, respectively. Otherwise we permuted 500,000
times. It took 79 hours to run all the permutations. If we ran at most
100,000 permutations, it took about 20 hours. In contrast, our method
only took 46 minutes. All the computation was done in a computing
server of Dual Xenon 2.4 Ghz.

\subsection{Comparing permutation $p$-values and nominal $p$-values}

The results we will report in this section are the property of
permutation $p$-values, instead of an artifact of our estimation
method. However, using direct permutation, it is infeasible to estimate
a very small permutation $p$-value, for example, $10^{-8}$ or less. In
contrast, our estimation method can accurately estimate such
permutation $p$-values efficiently.\footnote{Our method cannot estimate
those extremely small permutation $p$-values such as $10^{-20}$
reliably. This is simply because only a few genotype profiles can yield
such significant results even in the whole genotype space.
Nevertheless, those results correspond to unambiguously significant
findings even after Bonferroni correction. Therefore, permutation may
not be needed. See the Supplementary Materials [\citet{Sun09}]
for more
details.} This enables a study of the relationship between permutation
$p$-values and nominal $p$-values. Such a relationship can provide
important guidance for the sample size or power of a new study.

\begin{figure}

\includegraphics{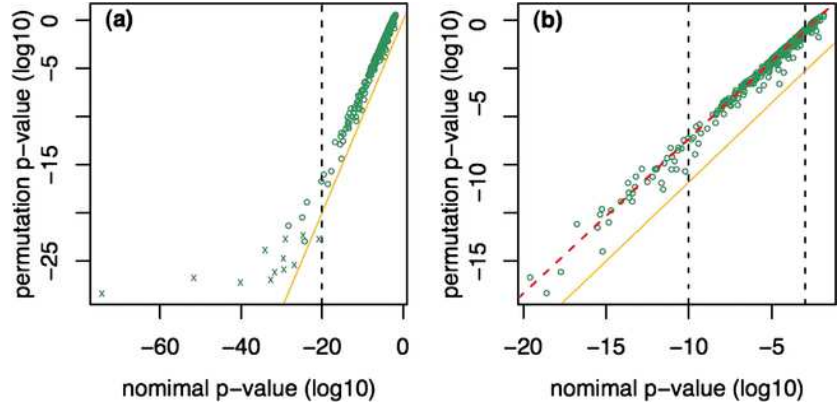}

\caption{Comparison of permutation $p$-value estimates and nominal
$p$-values. \textup{(a)} Scatter plot of permutation $p$-value
estimates vs. nominal $p$-value in log10 scale for the 500 gene
expression traits. Those unreliable permutation $p$-value estimates are
indicated by ``$x$.'' See footnote 2 for explanation. \textup{(b)}
Scatter plot for 483 gene expression traits with nominal $p$-value
larger than $10^{-20}$. In both \textup{(a)} and \textup{(b)} the solid
line is $y = x$. In \textup{(b)}, the broken line fitting the data is
obtained by median regression for those 359 genes with nominal
$p$-values between $10^{-10}$ and $10^{-3}$.}\label{fig4}
\end{figure}

Let $x$ and $y$ be $\log_{10}$(nominal $p$-value) and
$\log_{10}$(permutation $p$-value estimate) respectively. We compared
$x$ and $y$ across the randomly selected 500 gene expression traits
used in the previous section [Figure \ref{fig4}(a)] and found an
approximate linear relation.

%

We employed median regression (R function rq) to capture the linear
pattern [Figure \ref{fig4}(b)].\footnote{Most genes whose fitted values
differ from the observed values more than 2-folds are below the linear
patterns. These genes often have more outliers than other genes, which
may violate the $t$-test assumptions and bring bias to nominal
$p$-values.} If the nominal $p$-value was too large or too small, the
permutation $p$-value estimate might be inaccurate. Thus, we used the
359 gene expression traits with nominal $p$-value between $10^{-10}$
and $10^{-3}$ to fit the linear pattern (in fact, using all the 483
gene expression traits with nominal $p$-values larger than $10^{-20}$
yielded similar results, data not shown). The fitted linear relation is
$y = 2.52 + 0.978 x$. Note $x$ and $y$ are in log scale. In terms of
the $p$-values, the relation is $q = \eta p^\kappa= 327.5 p^{0.978}$,
where $p$ and $q$ indicate nominal $p$-value and permutation $p$-value,
respectively. If $\kappa=1$, ${q} = \eta p$, and $\eta$ can be
interpreted as the effective number of independent tests (or the
effective number of independent genotype profiles). However, the
observation that $\kappa$ is close to but smaller than 1 (lower bound
0.960, upper bound 0.985) implies that the effective number of
independent tests, which can be approximated by $q/p = \eta
p^{\kappa-1} = \eta p^{-0.022}$, varies according to the nominal
$p$-value $p$. For example, for $p = 10^{-3}$ and $10^{-6}$, the
expected effective number of independent tests is approximately 381 and
444, respectively.

The relation between the effective number of independent tests and the
significance level can be explained by the geometric interpretation of
permutation \mbox{$p$-values}. Given a nominal $p$-value cutoff, whether two
genotype profiles correspond to two independent tests amounts to
whether they can be covered by the same significance set. As the
$p$-value cutoff becomes smaller, the significance set becomes smaller
and, thus, the chance that two genotype profiles belong to one
significance set is smaller. Therefore, smaller $p$-value cutoff
corresponds to more independent tests.

\section{Discussion}\label{sec4}

In this paper we have proposed a geometric interpretation of
permutation $p$-values and a method to estimate permutation $p$-values
based on this interpretation. Both theoretical and empirical results
show that our method can estimate permutation $p$-values reliably,
except for those extremely small or relatively large ones. The
extremely small permutation $p$-values correspond to even smaller
nominal $p$-values, for example, $10^{-20}$. They indicate significant
linkages/associations even after Bonferroni correction; therefore,
permutation $p$-value evaluation is not needed. The relatively large
permutation $p$-values, for example, those larger than 0.1, can be
estimated by a small number of permutations, although in practice such
nonsignificant cases may be of little interest. The major computational
advantage of our method is that the computational time is constant
regardless of the significance level. This computational advantage
enables a study of the relation between nominal $p$-values and
permutation $p$-values in a wide range. We find that the effective
number of independent tests is not a constant; it increases as the
nominal $p$-value cutoff becomes smaller. This interesting observation
can be explained by the geometric interpretation of permutation
$p$-values and can provide important guidance in designing new studies.

Parallel computation is often used to improve the computational
efficiency by distributing computation to multiple
processors/computers. Both direct permutation and our estimation method
can be implemented for parallel computation. In the studies involving a
large number of traits (e.g., eQTL studies), one can simply distribute
an equal number of traits to each processor. If there are only one or a
few traits of interest, for direct permutation, one can distribute an
equal number of permutations to each processor. For our estimation
method, the most computationally demanding part (which takes more than
80\% of the computational time) is to estimate $P(r,\alpha)$, which can
be paralleled by estimating $P(r,\alpha)$ for different $r$'s
separately. Furthermore, for a particular $r$, $P(r,\alpha)$ is
estimated by evaluating the nominal $p$-values for a large number of
genotype profiles whose distances to the best partition are $r$. The
computation can be further paralleled by evaluating nominal $p$-values
for a subset of such genotype profiles in each processor.

As we mentioned at the beginning of this paper, we focus on the genetic
studies with high density markers, where the test statistics are
evaluated on each of the genetic markers directly. Our permutation
$p$-value estimation method cannot be directly applied to interval
mapping [\citet{Lander89}, \citet{Zeng93}]. However, we
believe that as the
expense of SNP genotype array decreases, most genetic studies will
utilize high density SNP arrays. In such situations, the interval
mapping may be no longer necessary.

We have discussed how to estimate the permutation $p$-value of the most
significant linkage/association. Permutation $p$-values can also be
used to assess the significance of each locus in multiple loci mapping.
\citet{Doerge96} have proposed two permutation-based thresholds for
multiple loci mapping, namely, the conditional empirical threshold
(CET) and residual empirical threshold (RET). Suppose $k$
markers have been included in the genetic model, and we want to test
the significance of the ($k+1$)th marker by permutation. The
samples can be stratified into $2^k$ genotype classes based on the
genotype of the $k$ markers that are already in the model (here
we still assume genotype is a binary variable). CET is evaluated based
on permutations within each genotype class. Alternatively, the
residuals of the $k$-marker model can be used to test the
significance of the ($k+1$)th marker. RET is calculated by
permuting the residuals across the individuals. RET is more powerful
than CET when the genetic model is correct since the permutations in
RET are not restricted by the $2^k$ stratifications. Our permutation
$p$-value estimation method can be applied to RET estimation without
any modification, and it can also be used to estimate CET with some
minor modifications. Specifically, let \textit{conditional desired
partitions} be the desired partitions that can be generated by the
conditional permutations. Then in equation (\ref{eq:05}), $N_{p} $
should be calculated as the number of conditional desired partitions
instead of the total number of desired partitions. In equation (\ref
{eq:06}), $P(r,\alpha)$ remains the same and $C_{U} (r)$ needs to be
calculated by counting the number of conditional desired partitions
within distance $r$ from at least one of the observed genotype
profiles.

There are some limitations in the current implementation of our method,
which are also the directions of our future developments. First, we
only discuss binary markers in this paper. The counting procedures in
Propositions 4 and 5 (see Section~IV in the Supplementary Materials
[\citet{Sun09}]) can be extended in a straightforward way to apply
to the genotypes with three levels. However, some practical
considerations need to be addressed carefully, for example, the
definition of the distance between genotype profiles and the choice of
the best partition. Second, the serial counting approximation relies on
the assumption that the correlated genotype profiles are close to each
other. This is true for genotype data in linkage studies, but in
general is not true for association studies, where the proximity of
correlated markers in haplotype blocks may be too coarse for immediate
use. We are investigating a clustering algorithm to reorder the
genotype profiles according to correlation rather than physical
proximity. Finally, our work here points toward extensions to the use
of continuous covariates, which can be applied, for example, to map
gene expression traits to the raw measurements of copy number
variations [\citet{Stranger07}].

\section*{Acknowledgments}

We appreciate the constructive and insightful comments from the editors
and the anonymous reviewers, which significantly improved this paper.
We acknowledge funding from EPA RD833825. However, the research
described in this article was not subjected to the Agency's peer review
and policy review and therefore does not necessarily reflect the views
of the Agency and no official endorsement should be inferred.

\begin{supplement}[id=suppA]
\stitle{Supplementary Methods and Results for ``A geometric
interpretation of the permutation \textit{p}-value and its application in eQTL
studies''}
\slink[doi]{10.1214/09-AOAS298SUPP}
\slink[url]{http://lib.stat.cmu.edu/aoas/298/supplement.pdf}
\sdatatype{.pdf}
\sdescription{The Supplementary Methods and Results include four
sections: (1) Single marker analysis and the choice of ``best
partition,'' (2) Description of genotype data, (3) Justification of the
hypersphere assumption for the balanced binary trait, and (4)
Propositions and the proofs.}
\end{supplement}

\printaddresses

\end{document}